# Thermally Controllable Multiple High Harmonics Generation by Phase-Change Materials-Mediated Fano Clusters

Arash Ahmadivand,[†‡]* Burak Gerislioglu,[†‡] and Nezih Pala[‡]

[‡]Department of Electrical and Computer Engineering, Florida International University, 10555 W Flagler St, Miami, Florida 33174, United States

*Corresponding Author: aahma011@fiu.edu

**ABSTRACT:** Substantial enhancement of nonlinear high-order harmonics generation based on Fano-resonant nanostructures has received growing interest due to their promising potential for developing integrated and advanced next-generation nanophotonic devices. In this study, going beyond conventional subwavelength structures, we propose an optothermally functional hetero-metallodielectric asymmetric eight-member octamer cluster composed of a central silicon nanodisk and peripheral disks with a phase-change material ($Ge_2Sb_2Te_5$). Using full electromagnetic calculations, we show that in the amorphous phase of the surrounding particles, the oligomer acts as an all-dielectric cluster, while in the crystalline regime, the octamer turns into a hybrid metallodielectric assembly. Exploiting the exquisite ability of supporting distinct Fano lineshapes at different wavelengths depending on the phase of $Ge_2Sb_2Te_5$, we showed the generation of both second and third harmonics at amorphous and crystalline phases of GST nanodisks, respectively with the produced harmonic wavelengths of 425 nm and 317 nm, respectively. Our calculations for the corresponding conversion efficiencies revealed significant enhancements as $\eta_{SHG}$=0.012% and $\eta_{THG}$=0.0081% for SHG and THG, respectively. Such an exquisite feature of multiresonant optothermally tunable cluster allows generation of several higher-harmonics with distinct intensities using a single system for future photonics applications.

High-harmonic generation (HHG) is a conventional phenomenon, has been observed initially in gas atoms (i.e. Ar, Kr, and Xe) due to multiple multiphoton ionization, resulting from the absorption of large number of incoming photons *via* nonlinear processes.[1-3] By producing wide variety of wavelengths from a single optical source, nonlinear HHG provides broad range of applications in modern nanophotonics technology and attosecond optics.[3-10] So far, several strategies have been developed to convert the fundamental incident beam frequency into the intense higher-order harmonics by subwavelength bulk solids with large field-induced of susceptibilities ($\chi^{(2)}$ or $\chi^{(3)}$) including but not limited to plasmonic antennas and waveguides,[11-15] all-dielectric structures,[16-21] photonic crystals,[22,23] ring resonators,[24-26] 2D and chiral materials,[27-30] etc. On the other hand, recently, intense and efficient ($P_{2\omega,3\omega}/P_\omega^2$) nonlinear spectral behavior of higher-harmonics have been reported in both Fano-resonant fully metallic, all-dielectric, and metallodielectric nanoparticle clusters with either simple or complex geometries, taking the advantage of dark side of plasmons.[15,17,19] It is well-accepted that efficient scattering of HHG signal into the far-field radiation can be realized by breaking the centrosymmetry of nanoparticle assemblies, resulting in strong electromagnetic field localization and formation of hotspots in nanosystems.[31] Although all of the elucidated mechanisms provide pronounced intensities and high efficiencies for generation of both odd and even harmonics, they suffer from limited tunability and high nonradiative and scattering losses.[32,33] It is shown that introducing low-loss and electrically controllable atomically-thin graphene layer to the subwavelength systems result in formation of higher-harmonics with enhanced tunability *via* tuning the doping concentration of graphene sheet.[34-37] However, manipulation and integration of atomically-thin layers with bulk systems require complex nanofabrication techniques, costly processes and also suffer from limited mechanical flexibilities. Moreover, in these systems, due to the weak field enhancement at the fundamental frequency, the near-field intensity in the higher-harmonics cannot be significantly boosted.[38] Consequently, finding an approach to control the nonlinear spectral response with high-harmonic intensity and functionality allows for developing advanced, tunable, and integrated nanophotonic devices. This can be realized by using optically controllable materials in the geometry of structure like thermally controllable phase-change materials (PCMs)[39,40] and semiconductors (e.g. InSb).[41,42] The latter option needs for external heating/cooling system to tune the corresponding intrinsic carrier density, which is limiting its efficacy in compact photonic systems. On the other hand, as new members of optical materials, optothermally controllable PCMs based on chalcogenide compounds offer novel and promising methods to address the inherent lack of tunability in conventional bulk solids.[39] Recent progresses in advanced nanophotonics technology have witnessed broad utilization of PCMs in designing optical devices such as antennas,[43,44] rewritable data storages,[45] modulators,[40,46-48] and beam steering metamaterials.[49,50] Possessing significantly different dielectric functions at two different phases (amorphous and crystalline states) of PCMs (e.g. $VO_2$,[43,51,52] AgInSbTe,[53] $Ge_xSb_yTe_z$[40,47-49]) at room-temperature, enabled emerging of novel optical devices with exotic properties. This exquisite feature of PCMs becomes more interesting, when we analyze the switching timescale between opposite phases and their rapid reversibility, which is around few tenth of nanoseconds.[54] This interplay between phases can be realized by applying heat, optical or electrical pulses as external thermal stimuli.[39]

In this study, we show generation of multiple higher-harmonics using an eight-member asymmetric Fano-resonant hetero-metallodielectric octamer assembly composed of central dielectric (silicon) and peripheral PCM ($Ge_2Sb_2Te_5$ or GST) nanodisks. Using the opposite behavior of GST nanodisks at below and above its critical temperature (~477 °C),[55] we showed that, due to having inherent asymmetric geometry, the proposed nanoassembly can be effectively tailored to support distinguished Fano dips at different wavelengths depending on the phase of the GST particles. By adjusting the fundamental wavelength of Fano dips in two different states of GST, we efficiently generated both second- (SHG) and third harmonics (THG) at different temperatures with high intensity. Our full electromagnetic analyses showed that in the amorphous regime of GST (a-GST), the structure acts as a full dielectric octamer cluster and supports a Fano minimum around $\lambda \approx 875$ nm. In contrast, when the GST state switches to the crystalline phase (c-GST), the satellite nanoparticles act similar to the metallic components. This results in the formation of a hybrid metallodielectric cluster, enabled to sustain Fano mode at $\lambda \approx 1050$ nm. This functionality is exploited for developing a platform to generate higher-harmonics according to the phase of the PCM and the fundamental wavelength of Fano lineshapes.

**Results and discussion**

**Excitation of Fano resonances.** Figure 1a demonstrates the scattering spectra for the proposed hetero-metallodielectric octamer cluster under intense plane wave radiation for both phases of GST nanoparticles and also contains a schematic for the asymmetric compositional cluster as an inset. The peripheral nanoparticles are GST compound and the central one is silicon, deposited on a glass ($SiO_2$) substrate with the relative permittivity of $\varepsilon \approx 2.1$. The corresponding dimensions of the optimized octamer are set as follows: the diameter of central and surrounding disks are 190 nm and 128 nm, respectively, with the homogenous height of 60 nm, and gap distance of 20 nm between neighboring nanoparticles. Here, we used experimental permittivity values reported by Palik[56] and Shportko *et al*.[57] for silicon and GST, respectively. Specifically, for the effective permittivity of GST at crystallization level, we employed the effective-medium expression based on Lorentz-Lorenz theory.[47,48,58,59]

Focusing on the spectral response of the octamer, we first analyze the optical properties of the cluster consists of satellite a-GST nanodisks. In this regime, the surrounding nanodisks have dielectric properties and the entire cluster can be considered as an all-dielectric nanoassembly. Employing discrete dipole approximation (DDA),[60] previous studies have shown that both symmetric and asymmetric all-dielectric nanoparticle oligomers can be tailored to support pronounced Fano resonances across the visible to the near-infrared region (Vis-NIR), originating from the destructive interference of the resonant and non-resonant modes excited in central and peripheral nanoparticles, respectively.[61-63] In the current structure, in the all-dielectric limit and under the linear polarization beam illumination, the Fano dip is induced at $\lambda \approx 875$ nm, which is recognized as a fundamental resonance frequency for the all-dielectric oligomer. Figure 1b illustrates the *xy*-plane electric-field ($E_y$) plot for the electric dipole mode excitation and totally opposite resonant behavior and resonance mismatch between the central and satellite nanodisks, consistent with the DDA mechanism. Conversely, switching the phase of the surrounding nanodisks from a-GST to c-GST leads to the formation of a metallodielectric nanoplasmonic assembly with a central silicon nanodisk. In this regime, the excited charges strongly oscillate and couple between the peripheral nanodisks and the central nanodisk does not hold any plasmonic moment and does not affect the plasmonic response. The influence of the presence of dielectric nanoparticles in the proximity and touching regimes with the metallic nanoparticles in an assembly have been described previously by using plasmon transmutation concept.[64,65] It is verified that insertion of dielectric nanoparticles in specific parts of a plasmonic assembly gives rise to the formation of collective magnetic modes as antibonding moments, enhancing the quality of the Fano lineshape. Similarly, in the scattering spectra, we monitored a weak magnetic dipole supported by the central nanodisk. On the other hand, in the scattering profile, a broad and net dipole moment as a bonding mode appeared along $\lambda$~900 to 1000 nm, while a narrow antibonding moment appeared at $\lambda \approx 1040$ nm with an asymmetric lineshape in between at $\lambda \approx 1015$ nm as the expected Fano minimum. Figures 1b and 1c exhibit the charge distribution maps (for $E_y$-field) across the metallodielectric cluster, computed using finite-element method (FEM) at the Fano dip wavelengths for both a-GST and c-GST clusters under *y*-polarized beam illumination to provide better understanding of the underlying the induced resonant modes of the assembly. For the all-dielectric octamer at the Fano dip position (Figure 1b), the central bigger nanodisk is non-resonant, while the peripheral ones support strong magnetic dipole moments. For the metallodielectric octamer at the Fano resonance wavelength (Figure 1c), we observed an antialignment in the excited resonant moments in peripheral nanodisks and also weak dipole moment in the central silicon disk. In addition, a weak coupling happened between the upper and lower nanodisks with the central one to complete the oscillation of magnetic charges for the excitation of Fano lineshape. However, comparing to the full metallic octamers,[67,68] in the currently analyzed cluster, substitution of central nanodisk with a silicon particle leads to slight reduction in the Fano dip depth and narrowness. On the other hand, it should be noted that due to the antisymmetric nature of octamer assembly in both amorphous and crystalline phases of the GST compound, it shows strong dependency on the incident polarization direction[67,68] which is not of interest here and we only analyze the spectral response for the *y*-polarized light illumination. Figures 1d and 1e demonstrate field enhancement and localization of plasmons (local E-field maps) in all-dielectric and metallodielectric octamers, respectively, at the Fano interference wavelength obtained by employing finite-difference time domain (FDTD) analysis, which are in complete agreement with the FEM studies. It is important to note that in the high-index dielectric nanoparticles, the particles are resonant, while in the metallic components, formation of hotspots between the particles and the localization of hot electrons is dominant. Defining the fundamental wavelengths of the structure in both examined regimes helps to adjust the incident time-domain beam source to generate nonlinear optical responses (see Methods).

**Multiple HHG.** Next, by considering the obtained spectral response for the proposed hetero-metallodielectric octamer, we study the feasibility of HHG at multiple wavelengths. To this end, by using the proposed method by Zhang *et al.*,[69] we only applied the surface tensor susceptibility component (normalized to the surface of the octamer) as a constant number for both $\chi^{(2)}_{s,nnn}, \chi^{(3)}_{s,nnn}$,[70] in our computations. This assumption is based on the fact that the dominant contribution in generation of higher harmonics is from the nanoclusters.[15] Figures 2a and 2b exhibit the SHG and THG using all-dielectric and metallodielectric octamers, respectively and show substantial enhancement in the near-field intensity at the Fano dip position. For the all-dielectric cluster and in the SHG limit, for the fundamental Fano resonance wavelength of $\lambda \approx 875$ nm ($\omega$), we observed formation of intense resonant mode around $\lambda \sim 425$ nm ($2\omega$). It should be underlined that there is a mismatch between the theoretically expected and numerically obtained frequency shift around ~12 nm, due to neglecting the influence of second harmonic susceptibility ($\chi^{(2)}$) dispersion.[19] The corresponding quality-factor (*Q*-factor) of the produced second harmonic is quantified as $Q=130$, which is exceedingly sharper and narrower in comparison to the analogous high-index nanoparticles-based all-dielectric nanoclusters, utilized for HHG.[16,17,19,71] The $E_y$-field maps in Figure 2c at the resonant condition show the magnetic responses of the cluster under incident electromagnetic beam illumination at the fundamental and SHG wavelengths, verifying the multifold field enhancement (around 45 times) in the produced nonlinear harmonic. On the other hand, for the oligomer cluster with c-GST nanoparticles, Figure 2b shows an intense THG compared to the fundamental Fano dip wavelength ($\lambda \approx 1015$ nm ($\omega$)), arising at $\lambda \approx 317$ nm ($3\omega$). Similar to previous analysis for the octamer in a-GST regime, we monitored a mismatch between the fundamental Fano dip and the produced third harmonic around ~21 nm, and the corresponding $Q$-factor of the THG lineshape is computed $Q\approx 169$. The $E_y$-field plots in Figure 2d at the resonant condition show the substantial augmentation of electric field in the cluster under incident electromagnetic beam illumination at the fundamental and THG wavelengths, conforming the multifold enhancement (around 55 times) in the produced nonlinear harmonic. The physical mechanism behind the origin of THG process can be better clarified by analyzing the effect of classically induced nonlinear polarization in general plasmonic systems. Here for the incident electromagnetic field $\tilde{E}(\mathbf{r},\omega)$, and the nonlinear susceptibility tensor for both second- and third-order harmonic ($\chi^{(2)}$ and $\chi^{(3)}$), the induced nonlinear polarizability can be written as: $\tilde{P}_i(\mathbf{r},2\omega) = \varepsilon_0 \chi_i^{(2)} \tilde{E}_i^2(\mathbf{r},\omega)$ and $\tilde{P}_i(\mathbf{r},3\omega) = \varepsilon_0 \chi_i^{(3)} \tilde{E}_i^3(\mathbf{r},\omega)$, respectively. For both all-dielectric and hybrid metallodielectric clusters, because of possessing similar behavior of the various elements in the nonlinear third-order susceptibility tensor, the electrical component of the incident field becomes normal to the hetero-metallodielectric octamer surface and the electric component of the field plays a fundamental role in the excitation of higher-order polarizations.[71-73] This mechanism and the strong intensity of the THG verify that the induced even- and odd-order harmonics using the proposed assembly is due to the hybridized plasmonic resonant modes and the induced Fano lineshape.

In continue, we compare the nonlinear performance of both all-dielectric and metallodielectric systems for nonlinear HHG. To this end, we first quantified the effective high-order nonlinear susceptibility using the proposed method by Boyd,[78] as $\chi^{(2)}=0.9\times 10^{-9}$ m².V⁻² and $\chi^{(3)}=0.99\times 10^{-17}$ m².V⁻² for octamer cluster with nanoparticles in a-GST and c-GST states, respectively. The obtained nonlinear response shows substantial performance of the proposed nanoplatform in two opposite regimes with an ability to support both SHG and THG, simultaneously. Moreover, we also calculated and compared the conversion efficiency of the proposed nanosystem. For a given excitation power, the corresponding SHG and THG conversion efficiencies can be estimated by: $\eta_{SHG} = P(2\omega)/P(\omega)$ and $\eta_{THG} = P(3\omega)/P(\omega)$, respectively, defined nominally up to 0.012% and 0.0081% for SHG and THG, respectively. Comparing the conversion efficiency of the current system with the previously reported values for similar all-dielectric and plasmonic systems shows the exquisite behavior of the proposed hetero-metallodielectric nanostructure. In Table 1 we listed the conversion efficiencies for both SHG and THG nonlinear processes in recently reported several simple and complex nanoscale platforms, verifying the exceptional performance of the proposed system. This comparison shows that the tailored nanostructure has a potential to compete with more complex and advanced systems.

Finally, to show the input power dependency of the nonlinear response of both SHG and THG, we calculated and plotted the normalized HHG intensity as a function of varying excitation pump power. As it can be observed in Figure 3a, at low excitation power below 40 mW, both second and third-order harmonics are following similar path to increase with the third power of the input. However, the big difference between the all-dielectric and hybrid clusters emerges at higher excitation powers beyond 40 mW. In this regime, due to strong thermal effects from the high power illumination, slight deviation happens in the SHG and THG process. This effect is more obvious in the octamer in a-GST phase, where the SHG is dominant. In the c-GST regime, the cluster is already conductive, hence, the thermal power cannot destructively influence the THG intensity. It should be noted that the required thermal heat to reverse the phase of the GST particles to amorphous state is much higher than the excitation power, therefore, the effect of high pump power on the harmonic generation process is negligible. Figure 3b exhibits the conversion efficiency in logarithmic scale as a function of excitation power variations. The increasing conversion efficiency ($\eta$) with the excitation power increment stops and saturates at the pump power above 60 mW for both targeted HHG processes.

**Conclusions**

In conclusion, by providing a systematic study, we have demonstrated the efficient and strong generation of both even- and odd-order harmonics using optothermally functional hetero-metallodielectric assembly composed of dielectric and PCM nanoparticles. The employed nanoassembly allowed us to excite Fano lineshapes at different wavelengths depending on the phase of the GST nanodisks. Assuming the resonant wavelengths as the fundamental wavelengths for the incident time domain beam, we analyzed the feasibility of both SHG and THG in the UV to the visible band with high conversion efficiencies. An active interplay between the amorphous and crystalline phases of GST substance allows for designing multiresonant nonlinear optical systems with an ability to tune the output wavelength for a targeted application. Such a functionality of GST compound in a nonlinear application is being reported for the first time. We believe that the proposed tunable platform paves the way for new methods for future studies in the utilization of optothermally controllable substances and novel devices for nonlinear photonic applications.

**Methods**

Electromagnetic wave simulations of the proposed hetero-metallodielectric octamer cluster were performed using both finite-difference time-domain (FDTD, Lumerical 2017) and finite-element (FEM, COMSOL Multiphysics 5.2) methods. The boundaries were surrounded by perfectly matched layers (PMLs). The incoming light for the incident beam for crystallization was a broadband plane wave with the bandwidth of 400 nm-1600 nm, with the irradiation power of $P_0$=3.2 μW, beam fluence of 60 Jm$^{-2}$, pulse duration of 500 fs, and repetition of 10KHz. We also defined another light source with the duration of 0.9 ns and irradiation power of 5.5 mW to offer the required thermal energy for the amorphization process. The spatial grid sizes with minimum lateral size 1 nm was applied and the Courant stability was satisfied by setting the time steps in FDTD to $dt$~0.1 fs. Then, for the heptamer cluster with the distance of "$r$" form the incident radiation source, the light fluence is defined by: $F(r) = 2P_0 \exp(-2r/w^2)/\pi w^2 f_r$,[79] where $w$ is the waist of the Gaussian beam. These analyses were performed by using Gaussian beam source with the amplitude of 10$^9$ and standard pulse length of ~1 ps with the Fano-resonant fundamental wavelength. The charge density plots were extracted by applying RF module in FEM computations and the E-field plots were obtained using the proper monitor in FDTD analysis. In addition, for the generation of high-order harmonics, we employed a time domain-based plane wave with the fundamental frequency according to the Fano dip position and pulse length of 2000 fs and offset of 4000 fs. The overall simulation time was set to 10 ps.

**Acknowledgements**

This work is supported by Army Research Laboratory (ARL) Multiscale Multidisciplinary Modeling of Electronic Materials (MSME) Collaborative Research Alliance (CRA) (Grant No. W911NF-12-2-0023, Program Manager: Dr. Meredith L. Reed).

**Authors Contribution**

A. A. and B. G. has equal contribution on this work.

**Additional Information**

**Competing financial interest**: The authors declare no competing financial interest.

Figures Captions

**Figure. 1. (a)** Scattering cross-section spectrum of the hetero-metallodielectric octamer cluster in both a-GST and c-GST phases of the surrounding nanoparticles. The inset is the schematic representation of the octamer assembly. **(b), (c)** The $|E_y|$-field maps for resonant modes excitation at the fundamental wavelength of the Fano modes for a-GST and c-GST states of cluster, respectively, obtained by FEM analysis. **(d), (e)** The $|E_y|$-field maps for resonant modes excitation and localization at the fundamental wavelength of the Fano modes for a-GST and c-GST states of cluster, respectively, obtained by FDTD analysis.

**Figure. 2. (a), (b)** SHG and THG spectra of the all-dielectric and hybrid metallodielectric octamer, respectively. **(c), (d)** The E-field distribution in both all-dielectric and metallodielectric clusters, respectively at the fundamental and the produced higher-order harmonics wavelengths, obtained by FDTD simulations.

**Figure. 3. (a)** Normalized HHG intensity of the all-dielectric and metallodielectric nanoclusters as a function of excitation power. **(b)** The conversion efficiency as a function of pump power for both SHG and THG in logarithmic scale.

Figure 1

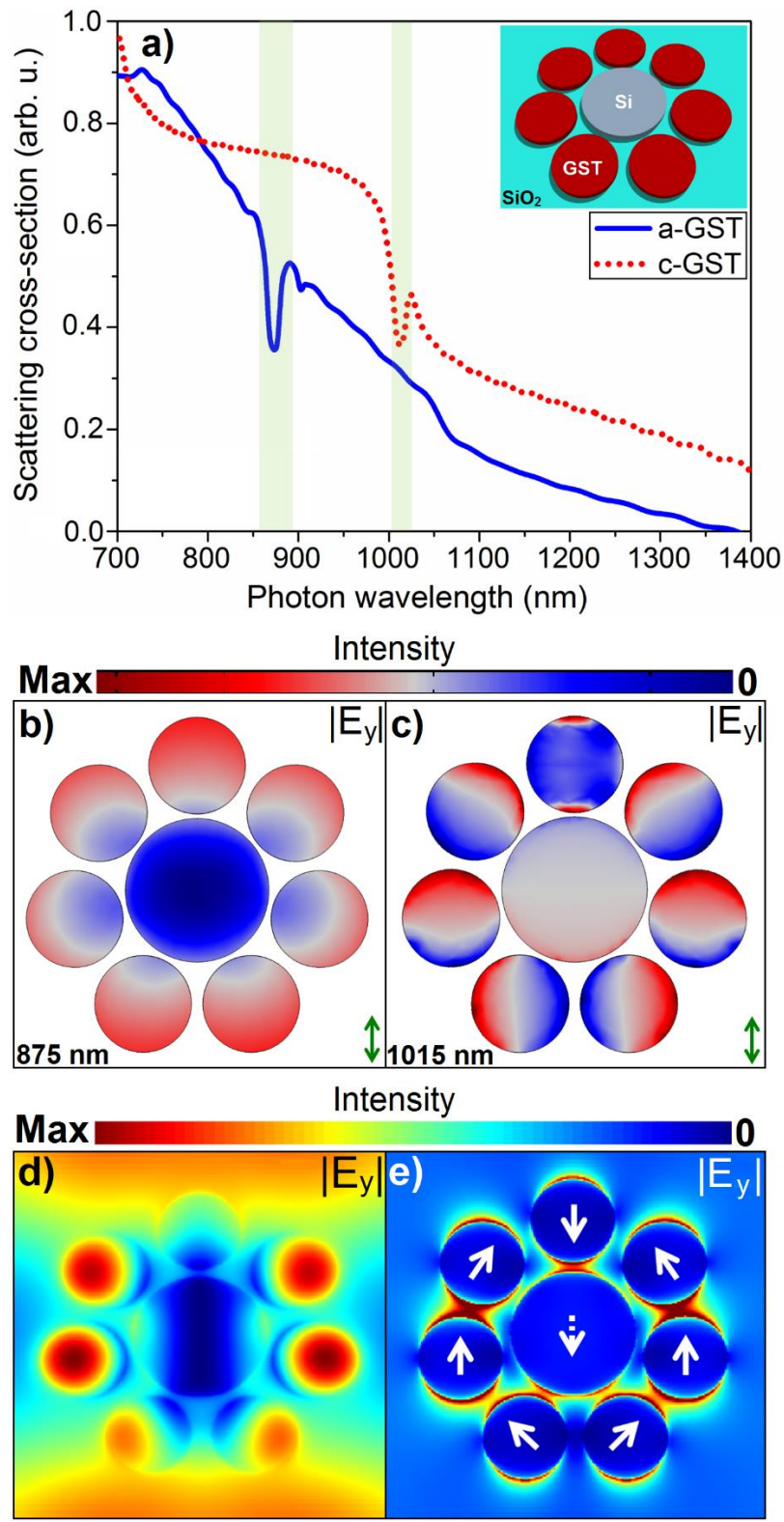

Figure 2

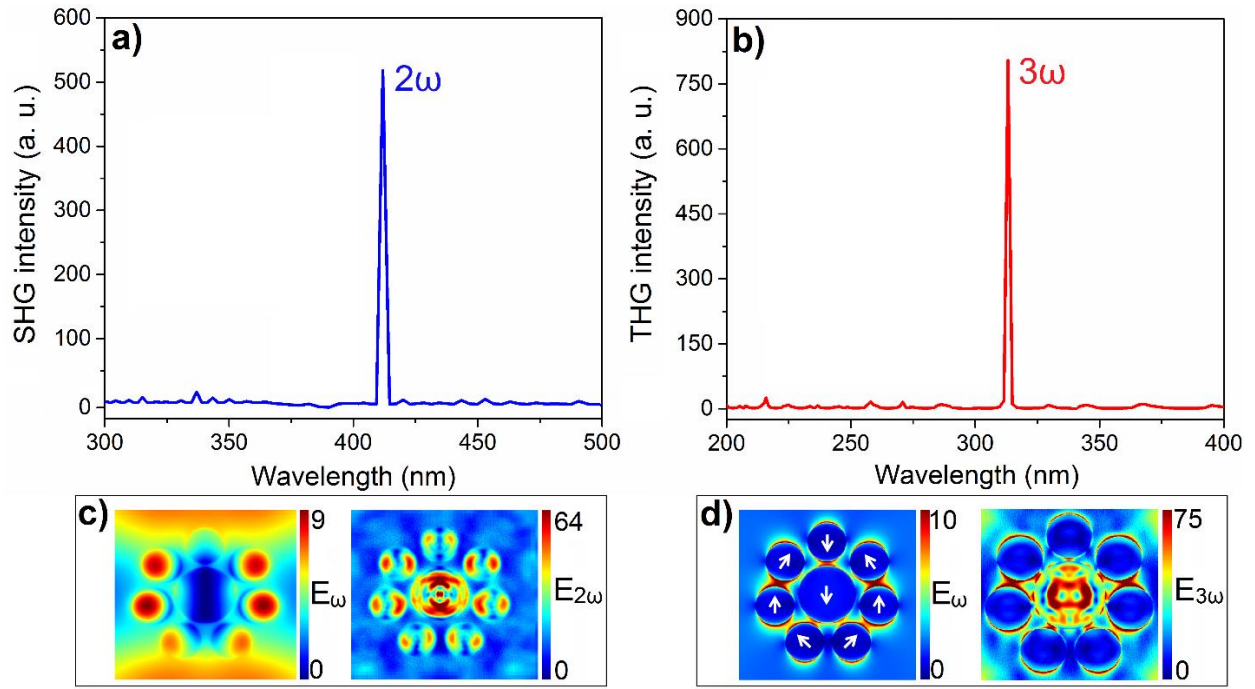

Figure 3

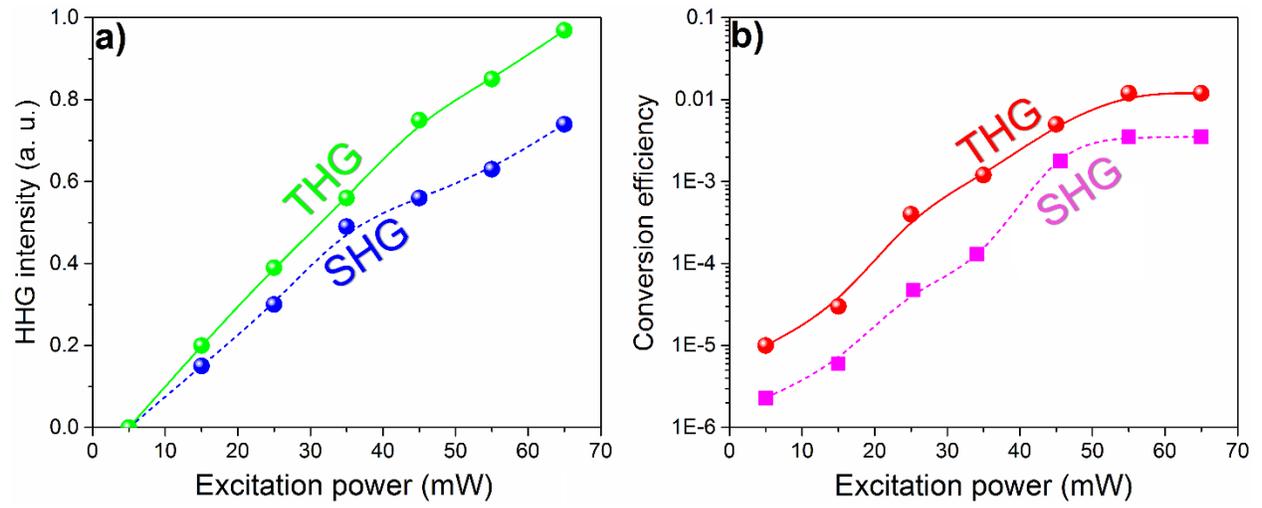

Table 1.

| Type of the high-order harmonic generation | Nonlinear systems | Efficiency ($\eta$) |
|---|---|---|
| SHG | 3D plasmonic metal-capped hemispherical nanoparticles | $1.8\times10^{-9}$ [69] |
| THG | Four-wave mixing and stimulated Raman scattering in a microcavity | $1.1\times10^{-9}$ [74] |
| SHG | Metamaterials coupled to quantum wells | $1.8\times10^{-11}$ [75] |
| THG | Single silicon nanodisks | $8\times10^{-8}$ [16] |
| THG | Single germanium nanodisks | $10^{-6}$ [18] |
| THG | Asymmetric plasmonic slot waveguides | $4.88\times10^{-6}$ [76] |
| THG | Metallodielectric core-shell hybrid nanoantenna | $7\times10^{-5}$ [77] |
| SHG | Current study | $8.1\times10^{-5}$ |
| THG | Current study | $1.2\times10^{-4}$ |